\documentclass[pra,twocolumn,showpacs,superscriptaddress,floatfix]{revtex4-1}
\usepackage{amsmath,amscd,amssymb,color}
\usepackage{graphicx,amsfonts,epsf}
\usepackage{epstopdf}
\usepackage{float}
\usepackage{hyperref}
\usepackage{enumerate}
\newcommand{\ie}{\textit{i.e.}}
\begin{document}
\title{Direct tomography of quantum states and processes via weak
measurements of Pauli spin operators on an NMR quantum processor}
\author{Akshay Gaikwad}
\email{ph16010@iisermohali.ac.in}
\affiliation{Department of Physical Sciences, Indian
Institute of Science Education \& 
Research Mohali, Sector 81 SAS Nagar, 
Manauli PO 140306 Punjab India.}
\author{Gayatri Singh}
\email{ph20015@iisermohali.ac.in}
\affiliation{Department of Physical Sciences, Indian
Institute of Science Education \& 
Research Mohali, Sector 81 SAS Nagar, 
Manauli PO 140306 Punjab India.}
\author{Kavita Dorai}
\email{kavita@iisermohali.ac.in}
\affiliation{Department of Physical Sciences, Indian
Institute of Science Education \& 
Research Mohali, Sector 81 SAS Nagar, 
Manauli PO 140306 Punjab India.}
\author{Arvind}
\email{arvind@iisermohali.ac.in}
\affiliation{Department of Physical Sciences, Indian
Institute of Science Education \& 
Research Mohali, Sector 81 SAS Nagar, 
Manauli PO 140306 Punjab India.}
\affiliation{Vice Chancellor, Punjabi University Patiala,
147002, Punjab, India}
%%%%%%%%%%%%%%%%%%%%%%%%%%%%%%%%%
\begin{abstract}
In this paper, we present an efficient weak measurement-based scheme for direct
quantum state tomography (DQST) and direct quantum process tomography (DQPT),
and experimentally implement it on an NMR ensemble quantum information processor
without involving any projective measurements. We develop a generalized quantum
circuit that enables us to directly measure selected
elements of the density matrix and
process matrix which characterize unknown quantum states and processes,
respectively.  This generalized quantum circuit uses the scalar $J$-coupling to
control the interaction strength between the system qubits and the metre qubit.
We experimentally implement  these weak measurement-based DQST and DQPT
protocols and use them to accurately characterize several two-qubit quantum
states and single-qubit quantum processes.  An extra qubit is used as a metre
qubit to implement the DQST protocol, while for the DQPT protocol, two extra
qubits (one as a metre qubit and the other as an ancilla qubit) are used.      
\end{abstract} 
\maketitle 
%%%%%%%%%%%%%%%%%%%%%%%%%%%%%%%
\section{Introduction}
\label{sec1}
In recent decades, the concepts of weak measurements and weak values have
attracted immense attention in quantum information processing from the
fundamental as well as the applications point of
view~\cite{yokota-njp-2009,avella-prl-2016}.  The weak value of a given
observable obtained via a weak measurement, although is in general a complex
number, has been shown to carry information about the system at times between
pre and post-selction~\cite{lev-prl-1990,lev-prl-1988}. Weak measurements  allow
us to sequentially measure incompatible observables so as to gain useful
information from the quantum system and learn about the initial state without
fully collapsing the state~\cite{tosi-prl-2016}.  This is in complete contrast
to conventional projective measurements, wherein the system collapses into one
of the eigenstates resulting in maximum state
disturbance~\cite{lundeen-prl-2016}.  This feature of weak measurements provides
an elegant way to address several important issues in quantum theory including
the reality of the wave function\cite{lundeen-nat-2011,bhati-pla-2022},
observation of a quantum Cheshire Cat in a matter-wave interferometer
experiment\cite{yuji-natcom-2014}, observing single photon trajectories in a
two-slit interferometer\cite{sacha-science-2011}, and the Leggett-Garg
inequality\cite{groen-prl-2013}. Weak measurements are also actively exploited
in the field of quantum information processing covering a wide range of
applications including quantum state and process
tomography\cite{lundeen-prl-2012,bolduc-natcom-2016}, state protection against
decoherence\cite{kim-np-2012,wang-pra-2014}, quantum state
manipulation\cite{blok-np-2014}, performing minimum disturbance
measurements\cite{baek-pra-2008}, precision measurements and quantum
metrology\cite{zhang-prl-2015}, sequential measurement of two non-commuting
observables\cite{avella-pra-2017} and tracking the precession of single nuclear
spins using weak measurements\cite{cujia-nature-2019}.

Several techniques have focused on direct estimation of quantum states and
processes including a method based on phase-shifting
technique~\cite{feng-pra-2021,li-pra-2022} and  direct measurement of quantum
states without using extra auxiliary states or post-selection
processes~\cite{wang-commun-2023}.  A selective QPT protocol based on quantum
2-design states was used to perform DQPT~\cite{perito-pra-2018} and
experimentally demonstrated on different physical
platforms\cite{paz-prl-2011,gaikwad-pra-2018,gaikwad-sr-2022}.  Conventional QST
and QPT methods require a full reconstruction of the density matrix and are
computationally resource intensive.  On the other hand, weak measurement based
tomography techniques have been used to perform state tomography and it was
shown that for certain special cases they outperform projective
measurements~\cite{debmalya,xu-prl-2021}.  An efficient DQST scheme was proposed
which directly measured arbitrary density matrix elements using only a single
strong or weak measurement~\cite{ren-prapp-2019}.  Circuit-based weak
measurement with post-selection has been reported on NMR ensemble quantum
information processor~\cite{lu-njp-2014}.

In this work, we propose an experimentally efficient scheme
to perform direct QST and QPT using weak measurements of
Pauli spin operators on an NMR ensemble quantum information
processor. The scheme allows  us to compute 
desired elements of the density matrix and is designed in such a way that
it does not require any ancillary qubits and has reduced
complexity as compared to recently proposed weak
measurement-based DQST and 
DQPT methods~\cite{kim-natcom-2018,zhang-ap-2017}.  Our scheme has three
major advantages, namely, (i) it  does not require
sequential weak measurements, (ii) it does not involve
implementation of complicated error-prone quantum gates such
as a multi-qubit, multi-control phase gate and (iii) it does
not require projective measurements.  Furthermore, our
proposed method is experimentally feasible as it requires a
single experiment to determine multiple selective elements
of the density/process matrix.  Our scheme is general and can be
applied to any circuit-based implementation. We
experimentally implemented the scheme to characterize several
two-qubit quantum states and single-qubit quantum processes
with high fidelity. Further, we fed the weak measurement
experimental results as input into a convex optimization
algorithm to reconstruct the underlying valid states and
processes\cite{gaikwad-qip-2021}.  We compared the experimentally
obtained state and process fidelities with theoretical
predictions and with numerical simulations and obtained a
good match within experimental uncertainties.

This paper is organized as follows: A brief review of weak
measurements and the detailed schemes for DQST and DQPT are
presented in Section~\ref{sec2}.  The details of the
experimental implementation of DQST and DQT via weak
measurements are given in Section~\ref{sec3}.
Section~\ref{sec3a} describes how to use an NMR quantum
processor to perform weak measurements of Pauli spin
operators, while Sections~\ref{sec3b} and \ref{sec3c}
contain details of a weak measurement of the Pauli operator
$\sigma_{1z}$ and the results of DQST and DQPT performed
using weak measurements, respectively.  Section~\ref{sec4}
contains a few concluding remarks.
%%%%%%%%%%%%%%%%%%%%%%%%%%%%%%%%%
\section{General scheme for direct QST and QPT via weak measurements}
\label{sec2}
%%%%%%%%%%%%%%%%%%%%%%%%%%%%%%%%%%%%%%%%%%%%%%%%%
\begin{figure*}  
{\includegraphics[scale=1.1]{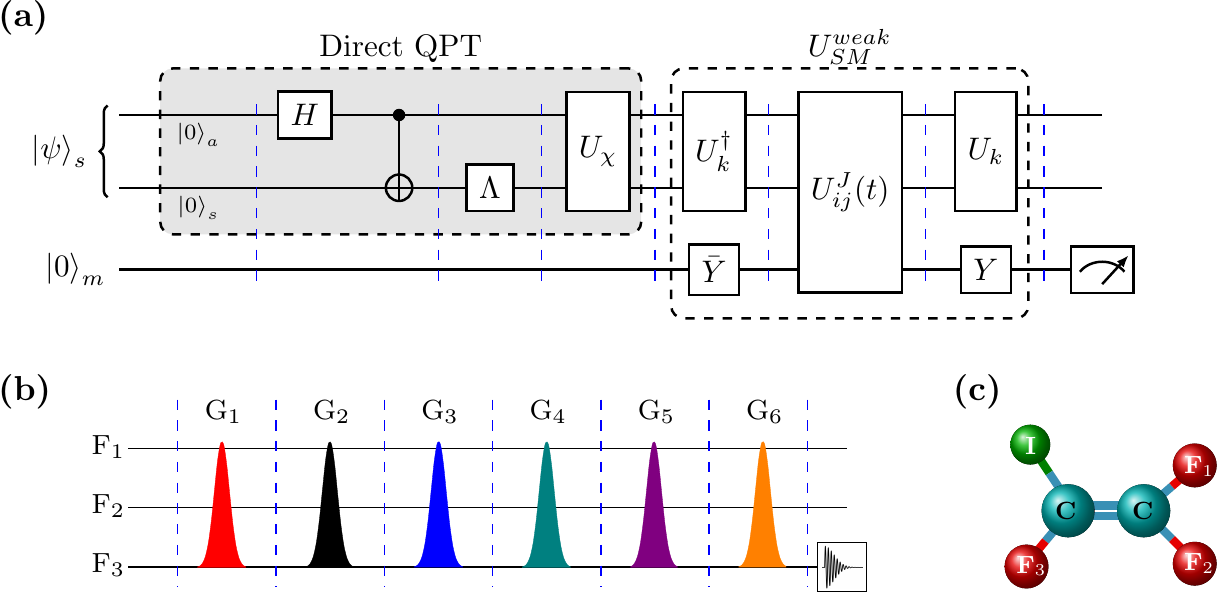}} 
\caption{(Color online)
(a)
General
quantum circuit for DQST of an initial unknown state $\vert
\psi \rangle_s$ and DQPT of a quantum channel $\Lambda$
using weak measurements.  The first (gray-shaded) block
corresponds to DQPT performed on the initial state $\vert 0
\rangle_a \vert 0 \rangle_s \vert 0 \rangle_m$. The unitary
operator $U_{\chi}$ depends on the operator basis in which
DQPT is performed.  The second (unfilled) block implements
the weak interaction between the system qubits and the meter
qubit, followed by measurement on the  meter qubit. 
(b) NMR implementation of 
the quantum circuit given in panel (a). 
The Gaussian-shaped curves represents GRAPE-optimized pulses 
($\rm{G}_i$) corresponding to given unitary operations on
all three qubits. (c) Structure of the molecule, trifluoroiodoethylene,
used to realize the three NMR qubits $\rm{F}_1$, $\rm{F}_2$ and $\rm{F}_3$.
}
\label{ckt} 
\end{figure*}
%%%%%%%%%%%%%%%%%%%%%%%%%%%%%%%%%%%%%%%%%%%%%%

Consider the system and the measuring device initially
prepared in a product state $\vert \psi \rangle 
\vert M \rangle$. The weak measurement
of an observable ${A}$ requires the evolution of the joint
state $\vert \psi, M \rangle$ under an operator of the form,
${U}_{SM} = e^{-ig {A}\otimes {B}} $, where $g$ is coupling
strength ($\vert g \vert \ll 1$) between the system and the
measuring device.  The operator ${B}$ corresponding to the
measuring device is chosen such that $\langle M \vert {B}
\vert M \rangle =0$.  In the weak measurement limit ($\vert
g \vert \ll 1$), the evolution operator can be approximated
upto first order in $g$ as, ${U}^{weak}_{SM} = ({I} -ig
{A}\otimes {B} ) $ and the evolution of the joint state
(system+measuring device) can be worked out as follows~\cite{lu-njp-2014}:
\begin{eqnarray}
\vert \psi, M \rangle_{\rm final}   & = &  e^{-ig {A}\otimes {B}}\vert \psi, M \rangle \nonumber  \\
 & \approx &  (I -ig {A}\otimes {B} ) \vert \psi, M \rangle \nonumber \\
 & = &  \vert \psi, M \rangle - ig {A} 
\vert \psi \rangle  \otimes {B} \vert M \rangle 
\label{eq1}
\end{eqnarray}
We will see that the above equation can be used for QST and
QPT by making appropriate measurements on the measuring
device.

Generally quantum processes are
either represented via the corresponding (i) $\chi$ matrix
(also referred to as the process matrix) using a Kraus
operator decomposition\cite{kraus-book-83} or (ii)
Choi-Jamiolkowski state using the channel-state duality
theorem~\cite{jiang-pra-2013}.  For an $N$-qubit system, the
$\chi$ matrix and the Choi-Jamiolkowski state corresponding
to quantum channel $\Lambda$ are given by~\cite{kraus-book-83,jiang-pra-2013}:
\begin{eqnarray}
\Lambda({\rho}_{in}) &=& \sum_{i=0}^{4^N-1} {K}_i
{\rho}_{in} K_i^{\dagger} = \sum_{m,n=0}^{4^N-1} \chi_{mn}
{E}_m \hat{\rho}_{in} {E}_n^{\dagger} \label{eq6} \\ 
\left|\Phi_{\Lambda}\right\rangle &=& (I \otimes \Lambda
)|\Phi\rangle=\frac{1}{2^{N / 2}} \sum_{m=0}^{2^{N}-1}| m
\rangle \otimes \Lambda| m \rangle  \label{eq7}
\end{eqnarray}
where $\chi_{mn}$ in Eq.~(\ref{eq6}) are elements of the $\chi$
matrix and $\vert \Phi_{\Lambda} \rangle$ in Eq.~(\ref{eq7})
is the Choi-Jamiolkowski state;  $\lbrace K_i \rbrace$'s and
the $\lbrace {E}_i \rbrace$'s in Eq.~(\ref{eq6}) are Kraus
operators and fixed basis operators respectively, while the
quantum state $\vert \Phi \rangle$ in Eq.~(\ref{eq7}) is a pure
maximally entangled state of $2N$ qubits $|\Phi\rangle=2^{-N
/ 2} \sum_{m=0}^{2^{N}-1}|m\rangle|m\rangle$. The density
matrix ${\rho}_{\Lambda} = \vert \Phi_{\Lambda} \rangle
\langle \Phi_{\Lambda} \vert $ corresponding to the
Choi-Jamiolkowski state can be mapped to the $\chi$ matrix
using an appropriate unitary transformation $U_{\chi}$ as
$\chi = U_{\chi} \rho_{\Lambda} U^{\dagger}_{\chi}$. The
unitary transformation matrix $ U_{\chi}$ depends only on
the fixed set of basis operators $\lbrace E_i\rbrace$
(Eq.~(\ref{eq6})) and does not depend on the quantum channel
to be tomographed. To perform DQPT of a given quantum
channel $\Lambda$ in terms of the $\chi$ matrix, we need to
apply the unitary transformation $ U_{\chi}$ on $\vert
\Phi_{\Lambda} \rangle$ and then follow the direct QST
protocol and estimate the desired elements $\chi_{mn}$.

Consider the operators $O^{\phi}_x = \vert \phi \rangle \langle \phi \vert
\otimes \sigma_x$ and $O^{\phi}_y = \vert \phi \rangle \langle \phi \vert
\otimes \sigma_y$ where $\vert \phi \rangle$ is a  pure
system state 
and $\sigma_{x(y)}$ are single-qubit Pauli spin operators. 
The expectation values of 
the $O^{\phi}_x$ and $O^{\phi}_y$ operators in
the weakly evolved joint state (Eq.~(\ref{eq1})) turn out to be:
\begin{eqnarray}
\langle O^{\phi}_x \rangle & = & ig \Big[ \langle \psi \vert
A^{\dagger} \vert \phi \rangle \langle \phi \vert \psi
\rangle - \langle \phi \vert A \vert \psi \rangle \langle
\psi \vert \phi \rangle \Big] \label{eq9}\\
\langle O^{\phi}_y \rangle & = & -g \Big[ \langle \psi \vert
A^{\dagger} \vert \phi \rangle \langle \phi \vert \psi
\rangle + \langle \phi \vert A \vert \psi \rangle \langle
\psi \vert \phi \rangle \Big] \label{eq10}
\end{eqnarray}
which can be simplified to:
\begin{equation}
\frac{\langle O^{\phi}_y \rangle - i \langle O^{\phi}_x
\rangle }{-2g} = 
\langle \phi \vert A \vert \psi \rangle
\langle \psi \vert \phi \rangle
\end{equation}
Straightforward algebra leads to 
\begin{equation}
\frac{\langle O^{\phi}_y \rangle - i \langle O^{\phi}_x
\rangle }{-2g} = \rho_{mn} = \langle m \vert \rho \vert n
\rangle \label{eq12}
\end{equation}
Similarly,
\begin{equation}
\frac{\langle O^{\phi}_y \rangle - i \langle O^{\phi}_x
\rangle }{-2g} =  \chi_{mn} = \langle m \vert \chi \vert n
\rangle  \label{eq13}
\end{equation}
Using Eqs.~(\ref{eq12}) and 
(\ref{eq13}), one can perform direct
QST and QPT by measuring $\langle O^{\phi}_x \rangle $ and
$ \langle O^{\phi}_y \rangle $ for an appropriate choice of
$A$ and $\vert \phi \rangle$.

It is interesting to note that 
\begin{equation}
\langle \phi \vert A \vert \psi \rangle
\langle \psi \vert \phi \rangle=\langle A
\rangle_{w}^{\phi}\Pi_{\psi}^{\phi}
\end{equation}
where $\langle A
\rangle_{w}^{\phi}$ is the weak value associated with a
post-selection of the system into the state $\phi$ and $
\Pi_{\psi}^{\phi}$ is the post-selection
probability~\cite{lundeen-prl-2016}
We do not use this connection is our work,
however, it connects our work with other schemes involving
weak values and post-selection.
%%%%%%%%%%%%%%%%%%%%%%%%%%%%%%%%%%%%%%%%
\section{NMR implementation of weak measurement scheme}
\label{sec3}
%%%%%%%%%%%%%%%%%%%%%%%%%%%%%%%%%%%%%%%%%%%%%%%%%
\begin{figure*}
\includegraphics[scale=1.05]{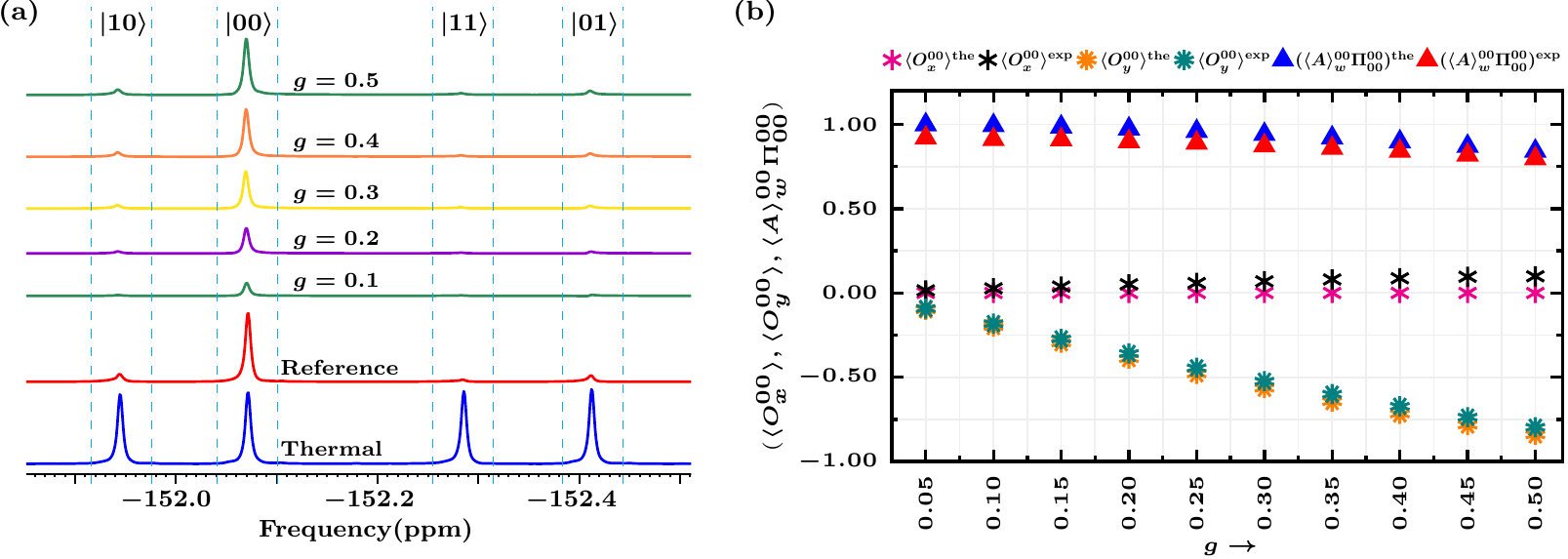}
\caption{(Color online) (a) NMR spectra obtained by measuring on the third
qubit ($\rm F_3$), corresponding to the meter qubit. The spectrum in blue
represent thermal equilibrium while the spectrum 
in blue represents the reference spectrum. The 
other spectra (from top) are obtained by implementing 
the quantum circuit for weak measurements
for different values of $g$, the
input state $\vert 00 \rangle_s \vert 0 \rangle_m$ and the weak interaction
unitary $U^{\rm weak}_{\rm SM}$ corresponding to 
the Pauli operator $\sigma_{1z}$ followed
by a $90^{\circ}$ phase shift on $\rm F_3$. (b) The theoretically and
experimentally obtained quantities $\langle O^{00}_x \rangle, \langle O^{00}_y
\rangle$ and $\langle A \rangle^{00}_w \Pi^{00}_{00}$ are compared 
for different
values of $g$. }
\label{fig1_zi}
\end{figure*}
%%%%%%%%%%%%%%%%%%%%%%%%%%%%%%%%%%%%%%%%%%%%%%
\subsection{NMR weak measurements of Pauli spin operators} 
\label{sec3a}
We used the three $\rm ^{19}F$ spins in the molecule trifluoroiodoethylene
dissolved in acetone-D6 to realize three qubits, denoting $\rm F_1$ and $\rm
F_2$ as the system qubits and $\rm F_3$  as the meter qubit (Fig.~\ref{ckt}(c)).
The rotating frame NMR Hamiltonian for a system of 
three spin-1/2 nuclei is given by~\cite{oliveira-book-07}:
\begin{equation}
\mathcal{H}=-\sum_{i=1}^{3} \nu_{i} I_{i z}+\sum_{i, j=1, i>j}^{3} J_{i j} I_{i z} I_{j z}
\end{equation}
where $\nu_i$ and $I_{iz}$ are the chemical shift and the $z$-component of the
spin angular momentum of the $i$th spin respectively, and $J_{ij}$ is the scalar
coupling between the $i$th and $j$th spins.  Experimental parameters
characterizing the given system can be found in the
Reference~\cite{gulati-epjd-2022}.

We set the initial state of the meter qubit to be $\vert M
\rangle = \vert 0 \rangle_m $, 
with $B = \sigma_x$  (Eq.~(\ref{eq1})). In this
case, the weak interaction evolution operator ${U}^{weak}_{SM}$ is of
the form:
\begin{equation}
{U}^{weak}_{SM} = {I} -ig {P_k}\otimes {\sigma_x} \label{eq15}
\end{equation}
where $I$ is an $8 \times 8$ identity matrix and $P_k = \lbrace I, \sigma_x,
\sigma_y, \sigma_z \rbrace^{\otimes 2} $ are two-qubit Pauli spin operators. 
The operator ${U}^{weak}_{SM}$ given in Eq.~(\ref{eq15}) can be decomposed as:
\begin{eqnarray}
\footnotesize	
{I} -ig {P_k}\otimes {\sigma_x} &=& {I} -ig (U_{k}\sigma_{iz} U_{k}^{\dagger}) \otimes (R_y(\frac{\pi}{2}) \sigma_z R_y^{\dagger}(\frac{\pi}{2})) \nonumber \\
 &=& \mathcal{U}_k ({I} -ig {\sigma_{iz}}\otimes {\sigma_z}) \mathcal{U}_k^{\dagger} \label{eq16}          
\end{eqnarray}
where $\sigma_{iz}$ is either $\sigma_{1z} = \sigma_z \otimes I$ or $\sigma_{2z}
= I \otimes\sigma_z $ and $\mathcal{U}_k = U_k \otimes R_y(\frac{\pi}{2})$;
$U_k$ is a two-qubit unitary operator acting on system qubits and is
constructed such that $P_k = U_{k}\sigma_{iz} U_{k}^{\dagger} $. To further
simplify Eq.~(\ref{eq16}), consider the $J$-evolution operator $U_{ij}^{J}(t)$:
\begin{equation}
U_{i j}^{J}(t)=e^{-i 2 \pi J_{ij} I_{i z} I_{j z} t }  
\label{eq17}
\end{equation}
If the evolution time $t$ is sufficiently small such that $g = \frac{\pi J_{ij}
t}{2} \ll 1$, Eq.~(\ref{eq17}) can be approximated as:
\begin{equation}
U_{i j}^{J}(t) \approx I - i g \sigma_{iz} \otimes \sigma_{jz}  \label{eq18}
\end{equation}
Hence using Eqs.~(\ref{eq16}) and (\ref{eq18}):
\begin{equation}
{U}^{weak}_{SM}  \approx  \mathcal{U}_{k} U_{ij}^J(t) 
\mathcal{U}_k^{\dagger} \label{eq19}
\end{equation}
where $t = \frac{2g}{\pi J_{ij}}$, $i = 1,2$ and $j=3$.

Hence the weak measurement of a desired Pauli operator $P_k$ can be performed by
applying the sequence of unitary operations given in Eq.~(\ref{eq19}) on an
initial joint state of the three-qubit system followed by the measurement of
$O^{\phi}_x$ and $O^{\phi}_y$. The list of $\mathcal{U}$s corresponding to all
$P_k$s is given in Table~\ref{table1}.

For the NMR implementation, the quantum circuit depicted in Fig.\ref{ckt}(a) is
divided into six parts, each consisting of a set of unitary operations.  Each of
the six parts are implemented using optimized pulse sequences generated through
GRAPE, which are represented graphically as colored Gaussian shapes in
Fig.~\ref{ckt}(b).  For example, the first part of the circuit consists of a
Hadamard gate followed by a CNOT gate, and this composite operation is
implemented using a GRAPE pulse denoted by $\rm{G}_1$ in Fig.\ref{ckt}(b).  The
approximate length of the GRAPE optimized rf pulses corresponding to given
quantum states (or processes) and Pauli operators is given in Table~\ref{grape}.
The power level was set to 28.57~W in all the experiments.

%%%%%%%%%%%%%%%%%%%%%%%%%%%%%%%%%%%%%%%%%%%%%
\begin{table}[h!]
\caption{Duration of GRAPE pulses (in ms) 
required to implement the weak measurement based scheme 
for operators $\sigma_{1z}$,
$\sigma_{1x}$, $\sigma_{2x}$ and $\sigma_{1x}\sigma_{2x}$ for 
a fixed value of
$g=0.2$. 
}
\setlength{\tabcolsep}{7pt} 
\renewcommand{\arraystretch}{1.4}
\footnotesize{
\begin{tabular}{c c c c c}%{lcdr} 
\hline \hline 
 &$\sigma_{1z}$&$\sigma_{1x}$&$\sigma_{2x}$&$\sigma_{1x}\sigma_{2x}$\\
 \hline \hline
 $\vert \psi_1 \rangle =(\vert 00 \rangle + \vert 11 \rangle)/\sqrt{2} $ & 5.3 & 5.1 & 5.2 & 5.4\\
 $\vert \psi_2 \rangle =(\vert 01 \rangle + \vert 10 \rangle)/\sqrt{2} $ & 5.3 & 5.2 & 5.3 & 5.6\\
 $\Lambda_1 = H $ & 2 & 0.5 & 1 & 1\\
 $\Lambda_2 = R_x(\frac{\pi}{2}) $ & 0.3  & 2.3 & 9.5 & 3.4 \\
 
\hline \hline 
\hline \end{tabular} }
\label{grape}
\end{table}
%%%%%%%%%%%%%%%%%%%%%%%%%%%%%%%%%%%%%%%% 

%%%%%%%%%%%%%%%%%%%%%%%%%%%%%%%%%%%%%%%%%%%%%%%%%%%%%%%%%%%%%%%
%%%%%%%%%%%%%%%%%%%%%%%%%%%%%%%%%%%%%%%%%%%%%
\begin{table}[h!]
\caption{Unitary operators $\mathcal{U}_k$ 
and $U_{i j}^{J}(t)$ required to implement 
the weak interaction operation ${U}^{{\rm weak}}_{SM}$ 
corresponding to all two-qubit Pauli operators $P_k$ }
\setlength{\tabcolsep}{12pt} % Default value: 6pt
\renewcommand{\arraystretch}{1.2}
\footnotesize{
\begin{tabular}{c c c}%{lcdr} 
\hline \hline $P_k$ &
~~~$\mathcal{U}_k$~~~& ~~~$U_{i j}^{J}(t)$~~~\\
\hline \hline $ \sigma_{2x} $ & $Y_2$ & $U_{23}^{J}(t)$   \\ 
$\sigma_{2y} $ & $\bar{X}_2 $ & $U_{23}^{J}(t)$  \\ 
$\sigma_{2z} $ & $I$ & $U_{23}^{J}(t)$  \\ 
$\sigma_{1x}$ & $Y_1 $ & $U_{13}^{J}(t)$  \\ 
$\sigma_{1x}\sigma_{2x} $ & $\bar{Z}_1 Y_2 U_{12}^{J}(\frac{1}{2J_{12}}) Y_1 $ & $U_{13}^{J}(t)$  \\ 
$\sigma_{1x}\sigma_{2y}$ & $\bar{Z}_1 \bar{X}_2 U_{12}^{J}(\frac{1}{2J_{12}}) Y_1 $ & $U_{13}^{J}(t)$  \\ 
$\sigma_{1x}\sigma_{2z}$ & $\bar{Z}_1 U_{12}^{J}(\frac{1}{2J_{12}}) Y_1 $ & $U_{13}^{J}(t)$ \\ 
$\sigma_{1y}$ & $\bar{X}_1 $ & $U_{13}^{J}(t)$  \\
$\sigma_{1y}\sigma_{2x}$ & $ Y_2 U_{12}^{J}(\frac{1}{2J_{12}}) Y_1 $ & $U_{13}^{J}(t)$  \\
$\sigma_{1y}\sigma_{2y}$ & $ \bar{X}_2 U_{12}^{J}(\frac{1}{2J_{12}}) Y_1 $ & $U_{13}^{J}(t)$   \\
$\sigma_{1y}\sigma_{2z}$ & $ U_{12}^{J}(\frac{1}{2J_{12}}) Y_1 $ & $U_{13}^{J}(t)$  \\
$\sigma_{1z}$ & $I$ & $U_{13}^{J}(t)$  \\
$\sigma_{1z}\sigma_{2x}$ & $X_1 Y_2 U_{12}^{J}(\frac{1}{2J_{12}}) Y_1 $ & $U_{13}^{J}(t)$  \\
$\sigma_{1z}\sigma_{2y}$ & $X_1 \bar{X}_2 U_{12}^{J}(\frac{1}{2J_{12}}) Y_1$ & $U_{13}^{J}(t)$ \\
$\sigma_{1z}\sigma_{2z}$ & $X_1 U_{12}^{J}(\frac{1}{2J_{12}}) Y_1 $ & $U_{13}^{J}(t)$ \\
\hline \end{tabular} }
\label{table1}
\end{table}
%%%%%%%%%%%%%%%%%%%%%%%%%%%%%%%%%%%%%%%% 
%%%%%%%%%%%%%%%%%%%%%%%%%%%%%%%%%%%%%%%% 

\subsubsection*{NMR Measurement of $O_x^{\phi}$ and $O_y^{\phi}$}
For simplicity, consider the post-selected state $\vert \phi \rangle$ to be one
of the computational basis vectors: $\lbrace \vert 00\rangle, \vert 01 \rangle,
\vert 10 \rangle, \vert 11 \rangle \rbrace$ which are required to
perform DQST or DQPT (Eq.~(\ref{rho})). In this case, it turns out that
the observables $O_{x(y)}^{\phi}$ can 
be directly measured by acquiring the NMR signal
from $\rm F_3$ (the meter qubit).  The NMR signal of $\rm F_3$
consists of four spectral peaks (see thermal spectra in blue color in
Fig.~\ref{fig1_zi}(a)) corresponding to four transitions 
associated with density matrix elements (referred to as readout 
elements):~$\rho_{56}$, $\rho_{12}$, $\rho_{78}$ and
$\rho_{34}$~\cite{gaikwad-qinp-2022}. 
The first peak from 
the left 
in Fig.\ref{fig1_zi}(a) 
corresponds to the post-selected state $\vert \phi \rangle = \vert 10
\rangle$ while 
the second, third and fourth peaks correspond to 
the post-selected states $\vert 00 \rangle$, $\vert
11 \rangle$ and $\vert 01 \rangle$, respectively. These peaks (from 
the left) are also
associated with readout elements $\rho_{56}$, $\rho_{12}$, $\rho_{78}$ and
$\rho_{34}$, respectively. 
The line intensity of the absorption mode spectrum ($x$-magnetization) 
is proportional to the real part of 
the corresponding readout element of the density matrix while the
dispersion mode spectrum ($y$-magnetization) is proportional 
to the imaginary part of the corresponding readout
element: 
\begin{equation}
 \langle O_{x}^{\phi} \rangle \propto {\rm Re}(\rho_{ij}) \quad {\rm and} \quad \langle O_{y}^{\phi} \rangle \propto {\rm Im}(\rho_{ij})
\end{equation}
where $\rho_{ij}$ is the readout element of the three-qubit density matrix on
which the observables $O_{x(y)}^{\phi}$ are being measured. The complete list of
observables $\langle O_{x(y)}^{\phi} \rangle$ with corresponding spectral
transitions and readout elements are listed in Table~\ref{table2}.
Note that in the case of an arbitrary post-selected state $ \vert \phi \rangle$,
one has to decompose the observables $O_{x(y)}^{\phi}$ into Pauli basis
operators as $O_{x(y)}^{\phi} = \sum_i a_i^{x(y)}P_i$, then measure $\langle P_i
\rangle$ corresponding to non-zero coefficients $a_i^{x(y)}$ and finally compute
$\langle O_{x(y)}^{\phi} \rangle$ for the given $\vert \phi \rangle$.  An
efficient way of measuring the expectation value of any Pauli observable is
described in Reference~\cite{singh-pra-2018}.

%%%%%%%%%%%%%%%%%%%%%%%%%%%%%%%%%%%%%%%%%%%%%
\begin{table}[h!]
\caption{Expectation values $\langle O_{x(y)}^{\phi} \rangle$ 
corresponding to spectral transitions and readout elements 
for the post-selected state $\vert \phi \rangle$ (which is
one of the computation basis vectors)}
\setlength{\tabcolsep}{4pt} 
\footnotesize{
\begin{tabular}{c c c c}%{lcdr} 
\hline \hline 
$\langle O_{x(y)}^{\phi} \rangle$ &~~~Transitions~~~&~~~elements~~~&~~~Peak (from left)~~~\\
\hline \hline 
$\langle O_{x(y)}^{00} \rangle$ & $\vert 000 \rangle \leftrightarrow \vert 001 \rangle$ & $\rho_{12}$ & second   \\ 
$\langle O_{x(y)}^{01} \rangle$ & $\vert 010 \rangle \leftrightarrow \vert 011 \rangle$ & $\rho_{34}$ & fourth  \\ 
$\langle O_{x(y)}^{10} \rangle$ & $\vert 100 \rangle \leftrightarrow \vert 101 \rangle$ & $\rho_{56}$ & first   \\ 
$\langle O_{x(y)}^{11} \rangle$ & $\vert 110 \rangle \leftrightarrow \vert 111 \rangle$ & $\rho_{78}$ &  third  \\ 
\hline \end{tabular} }
\label{table2}
\end{table}
%%%%%%%%%%%%%%%%%%%%%%%%%%%%%%%%%%%%%%%% 
%%%%%%%%%%%%%%%%%%%%%%%%%%%%%%%%%%%%%%%%%%%%%%%%%
\begin{figure}
{\includegraphics[scale=0.90]{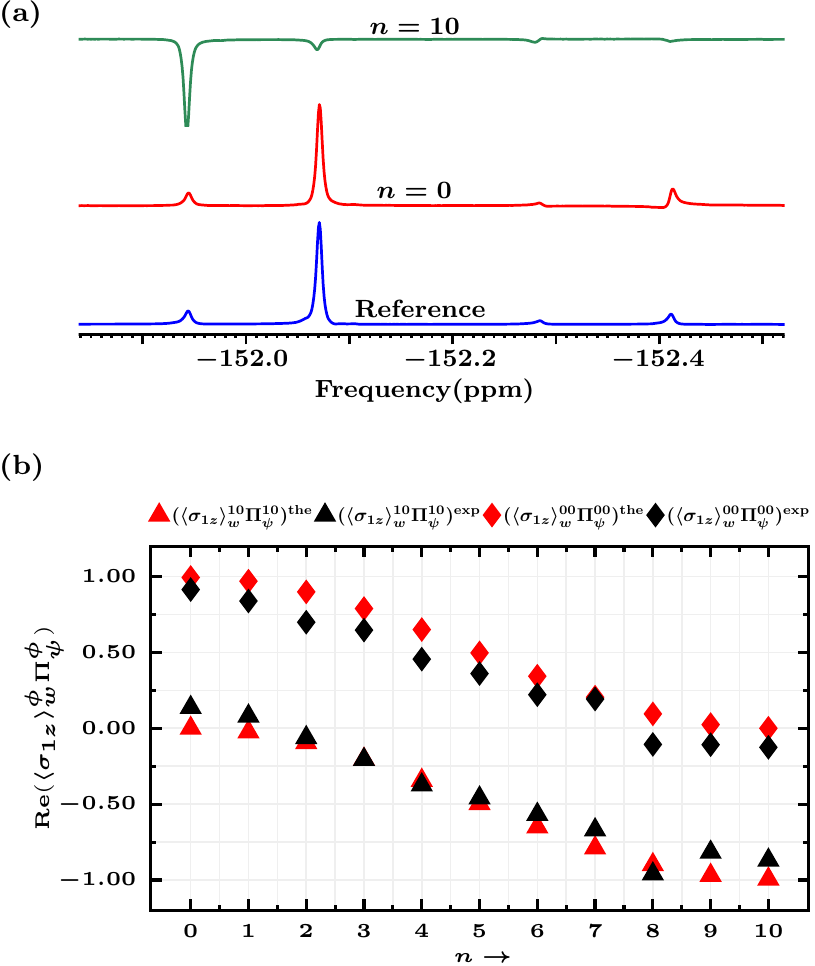}}
\caption{(Color online) (a) NMR spectra obtained after implementing the quantum
circuit for weak measurements on the initial state $\vert \psi \rangle_s =
\cos{(\frac{n \pi}{20})} \vert 00 \rangle + \sin{(\frac{n \pi}{20})} \vert 10
\rangle $ for $g = 0.1$ and the observable $\sigma_{1z}$. The spectra in green
and red correspond to $n=10$ and $n=0$, respectively.  (b) 
Plots comparing the experimentally measured
$\langle \sigma_{1z} \rangle_w^{\phi} \Pi_{\psi}^{\phi}$ with
its theoretical value as a function of 
the initial state $\vert \psi \rangle_s =
\cos{(\frac{n \pi}{20})} \vert 00 \rangle + \sin{(\frac{n \pi}{20})} 
\vert 10 \rangle$.  
}
\label{fig2_zi}
\end{figure}
%%%%%%%%%%%%%%%%%%%%%%%%%%%%%%%%%%%%%%%%%%%%%%
\subsection{Experimental weak measurement of $\sigma_{1z}$ }
\label{sec3b}
As an illustration, we experimentally obtained various relevant quantities (as
described in Section~\ref{sec2}) for the two-qubit Pauli operator
$\sigma_{z}\otimes I_{2 \times 2} = \sigma_{1z}$. The weak measurement of
$\sigma_{1z}$ allows us to measure all the diagonal elements of the density
matrix in a single experiment.  We experimentally implemented the proposed
scheme and measured $\langle O_{x(y)}^{\phi} \rangle$ and $A_w^{\phi}
\Pi_{\psi}^{\phi}$ with varying weak interaction strength $g$ for the case of $A
= P_k = \sigma_{1z}$, with the initial state $\vert \psi \rangle = \vert 00
\rangle$ and $\vert \phi \rangle$ being the computational basis vectors.  All
the NMR spectrums in Fig.\ref{fig1_zi}(a) correspond to measurements on the $\rm
F_3$ qubit. The bottom spectrum in blue represents the thermal equilibrium
spectrum obtained by applying a readout pulse on thermal state  followed by
detection on the $\rm F_3$ qubit. As shown in Table.\ref{table2}, the first peak
(from the left) in the thermal spectrum corresponds to $\langle O_{x(y)}^{10}
\rangle$, while the second, third and fourth  peaks correspond to $\langle
O_{x(y)}^{00} \rangle$, $\langle O_{x(y)}^{11} \rangle$ and $\langle
O_{x(y)}^{01} \rangle$, respectively. 

The reference spectrum depicted in red is obtained by applying a readout pulse
on an experimentally prepared pseudo pure state (PPS) using the spatial
averaging technique\cite{oliveira-book-07, cory-physicad-1998} followed by
detection on the $\rm F_3$ qubit.  The value of the reference peak is set to be
1. With respect to this reference, the observable $\langle O_{x}^{\phi} \rangle$
can be directly measured by computing spectral intensity by integrating the area
under the corresponding peak, while the observable $\langle O_{y}^{\phi}
\rangle$ can be measured by first performing a $90^{\circ}$ phase shift and then
computing the intensity. Note that the quantity $\langle O_{y}^{\phi} \rangle$
is (-1) times the spectral intensity. 

For example, the third spectrum (depicted in green) in Fig.\ref{fig1_zi}(a)
corresponds to $g = 0.1$ The peak intensity with respect to reference spectrum
turns out to be $0.1821 \pm 0.003$ which gives $\langle O_{y}^{00} \rangle =
-0.1821 \pm 0.003$ while the experimental value of $\langle O_{x}^{00} \rangle$
(intensity before $90^{\circ}$ phase shift) turns out to be $0.0272 \pm 0.0066$.
Similarly, the other four spectra in Fig.\ref{fig1_zi}(a) correspond to various
$g$ values.  One can see that from Fig.\ref{fig1_zi}(a), the for all values of
$g$ the spectral intensity of the first, third and fourth peak corresponding to
the post-selected states $\vert 10 \rangle$, $\vert 11 \rangle$ and $\vert 01
\rangle$, respectively, is negligible as compared to the reference peak which
implies that the quantities $\langle O_{x(y)}^{10} \rangle$, $\langle
O_{x(y)}^{11} \rangle$ and $\langle O_{x(y)}^{01} \rangle$ are almost zero.
This is to be expected since theoretically $\langle \phi \vert \sigma_{1z} \vert
\rho \vert \phi \rangle = 0$ except for $ \vert \phi \rangle = \vert 00 \rangle
$, whereas the spectral intensity of the second peak corresponding to $\langle
O_{y}^{00} \rangle$ is non-zero and increases with $g$. 

The experimental values $\langle O_{x(y)}^{00} \rangle$ and ${\langle
\sigma_{1z} \rangle}_w^{00} \Pi_{00}^{00}$ are compared with their theoretically
expected values in Fig.~\ref{fig1_zi}(b), for different values of $g$. Only the
real part of ${\langle \sigma_{1z} \rangle}_w^{00} \Pi_{00}^{00}$, \ie  ${\rm
Re}({\langle \sigma_{1z} \rangle}_w^{00} \Pi_{00}^{00}) = \frac{\langle
O_{y}^{00} \rangle}{-2g} $ is plotted as the imaginary part turns out to be
almost zero (as seen from $\langle O_{x}^{00} \rangle$ values).  The
experimental quantity ${\langle \sigma_{1z} \rangle}_w^{00} \Pi_{00}^{00}$ was
calculated using Eq.~(\ref{eq11}), however it can also be computed by rescaling
the spectrum by the factor $\vert \frac{1}{2g} \vert$ with respect to the
reference spectrum. The expected value of ${\langle \sigma_{1z} \rangle}_w^{00}
\Pi_{00}^{00}$ is equal to 1, which is the density matrix element $\rho_{11}$ of
initial state $\vert \psi \rangle = \vert 00 \rangle$. 

As the value of $g$ increases, the experimental and theoretical value of
$\rho_{11}$ deviates more and more from 1, because the weak interaction
approximation no longer holds for relatively large values of $g$. At $g = 0.05$,
the experimental value of $\rho_{11}^{\rm exp}$ was $(0.9198 \pm 0.0057) + i
(0.0525 \pm 0.0399)$, while at $g = 0.5$ the value of $\rho_{11}^{\rm exp}$ was
$(0.7971 \pm 0.0021) + i (0.0989 \pm 0.0439)$.  We also would like to point out
here that in real experiments an arbitrary small value of $g$ may not work,
since the signal strength after the weak interaction may be too small to detect
and may introduce large errors in the measurements.

We also implemented the weak measurement-based scheme for different initial
states.  The results shown in Fig.\ref{fig2_zi} were obtained by experimental
implementing the weak measurement-based scheme for a fixed interaction strength
$g = 0.1$ and for different initial states of the form $\vert \psi \rangle =
\cos{(\frac{n \pi}{20})}\vert 00 \rangle + \sin{(\frac{n \pi}{20})}\vert 10
\rangle$.  The  NMR spectrum in blue color in Fig.\ref{fig2_zi}(a), is the
reference spectrum, while the other two spectra in red and green correspond to
the states $n = 0$ and $n = 10$, respectively, which were obtained by
implementing the weak measurement quantum circuit (Fig.\ref{ckt}) followed by a
$90^{\circ}$ phase shift. Note that since the value of $g$ is fixed, the spectra
corresponding to all $n$ are rescaled by the factor $\frac{1}{2(0.1)} = 5$ with
respect to the reference spectrum, which directly yields ${\rm Re}( {\langle
\sigma_{1z} \rangle}_w^{\phi} \Pi_{\psi}^{\phi} )$ and $ {\rm Im}({\langle
\sigma_{1z} \rangle}_w^{\phi} \Pi_{\psi}^{\phi} )$ (the real and imaginary parts
of corresponding density matrix, respectively).  For $n = 0$ (red spectrum), the
observables $\langle O_{y}^{10} \rangle$, $\langle O_{y}^{00} \rangle$, $\langle
O_{y}^{11} \rangle$ and $\langle O_{y}^{01} \rangle$ turned out to be $-0.1370
\pm 0.0008$, $-0.9137 \pm 0.0038$, $-0.0267 \pm 0.0010$ and $-0.1243 \pm
0.0085$, respectively.  For $n = 10$ (green spectrum) the observables turned out
to be $0.8687 \pm 0.0054$, $0.1255 \pm 0.0033$, $0.0294 \pm 0.0003$ and $0.0237
\pm 0.0030$, respectively. 
The experimentally obtained
${\rm Re}\langle {\sigma_{1z}}_w^{\phi} \Pi_{\psi}^{\phi} \rangle$ 
is compared
with the theoretically 
expected values 
in Fig.~\ref{fig2_zi}(b), 
for $\vert \phi \rangle = \vert 00 \rangle$ and $\vert
\phi \rangle = \vert 10 \rangle$ and for various initial states.  
The experimental values are in very good 
agreement with the theoretical predictions in both
Figs.\ref{fig1_zi} and \ref{fig2_zi}, which clearly shows the successful
implementation of the weak measurement of $\sigma_{1z}$. 

%%%%%%%%%%%%%%%%%%%%%%%%%%%%%%%%%%%%%%%%%%%%%%%%%
\begin{figure}[t] {\includegraphics[scale=0.95]{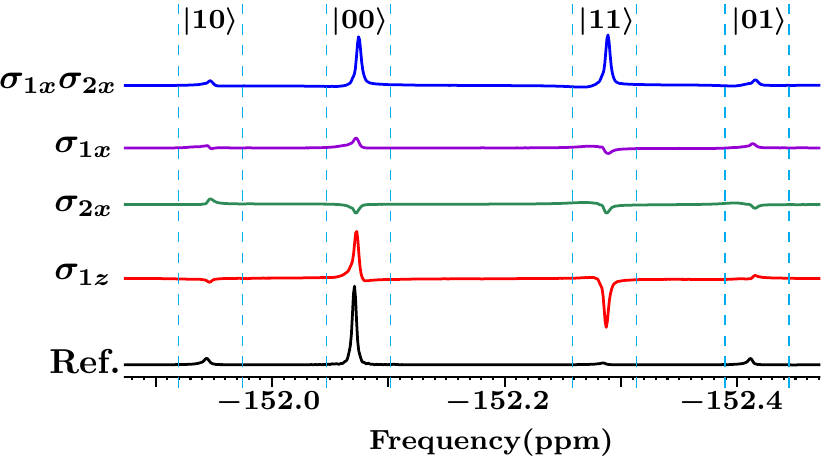}}
\caption{(Color online) Experimental readouts 
demonstrating DQST of the Bell state $\vert \psi
\rangle_s = \frac{1}{\sqrt 2} (\vert 00 \rangle + \vert 11 \rangle) $ 
using weak
measurements for 
a value of $g=0.2$. 
NMR spectra for
the observables $\sigma_{1z}$ (red),
$\sigma_{2x}$(green), 
$\sigma_{1x}$(purple), 
and $\sigma_{1x}\sigma_{2x}$
(blue)
were obtained by 
implementing the weak measurement quantum
circuit, followed by a $90^{\circ}$ phase shift on the initial
Bell state.}
\label{bell_spectra}
\end{figure}
%%%%%%%%%%%%%%%%%%%%%%%%%%%%%%%%%%%%%%%%%%%%%%
\subsection{Experimental DQST and DQPT using weak measurements}
\label{sec3c}
We now proceed to experimentally demonstrate element-wise full reconstruction of
the density and process matrices of several states and quantum gates using the
proposed weak measurement-based DQST and DQPT schemes.  To estimate a desired
element $\rho_{mn}$ of the density matrix or $\chi_{mn}$ of 
the process matrix, one of
the possible choices of the post-selected state $\vert \phi \rangle$, together
with the Pauli operator $P_k$, is depicted as $(\vert \phi \rangle, P_k)$ 
in the matrix:
\begin{equation}  \label{rho}
\footnotesize	
\begin{pmatrix}
(\vert 00 \rangle, \sigma_{1z}) & (\vert 01 \rangle, \sigma_{2x}) & (\vert 10 \rangle, \sigma_{1x}) & (\vert 11 \rangle, \sigma_{1x}\sigma_{2x}) \\
\rho_{12}^* & (\vert 01 \rangle, \sigma_{1z}) & (\vert 10 \rangle, \sigma_{1x}\sigma_{2x}) & (\vert 11 \rangle, \sigma_{1x}) \\
\rho_{13}^*  & \rho_{23}^* & -(\vert 10 \rangle, \sigma_{1z}) & (\vert 11 \rangle, \sigma_{2x}) \\
 \rho_{14}^* & \rho_{24}^* & \rho_{34}^* & -(\vert 11 \rangle, \sigma_{1z}) \\
\end{pmatrix}
\end{equation}
In this case, the full QST of a two-qubit quantum state requires weak
measurements of only four Pauli operators: $\lbrace  \sigma_{1z}, \sigma_{1x},
\sigma_{2x}, \sigma_{1x} \sigma_{2x} \rbrace$.  The weak measurement of
$\sigma_{1z}$ allows us to directly estimate all the diagonal elements
representing the populations of the energy eigenstates, while the weak
measurements of $\sigma_{1x}$, $\sigma_{2x}$ and $\sigma_{1x} \sigma_{2x}$ yield
two off-diagonal elements, each representing a single- and a multiple-quantum
coherence.

As  an illustration, we experimentally performed DQST of the maximally entangled
Bell states: $\vert \psi_1 \rangle = (\vert 00 \rangle + \vert 11
\rangle)/\sqrt{2} $ and $\vert \psi_2 \rangle = (\vert 01 \rangle + \vert 10
\rangle)/\sqrt{2} $ as well as DQPT of two quantum gates:  the Hadamard gate $H$
and a rotation gate $R_x(\frac{\pi}{2})$.  For both DQST and DQPT, the value of
$g$ is set to be $0.2$. 

%%%%%%%%%%%%%%%%%%%%%%%%%%%%%%%%%%%%%%%%%%%%%
\begin{table}[h!]
\caption{Experimental state ($\vert \psi \rangle$) 
and process($\Lambda$) fidelities 
obtained using weak measuremet-based DQST and DQPT}
\setlength{\tabcolsep}{7pt} % Default value: 6pt
\renewcommand{\arraystretch}{1.6} % Default value: 1
\footnotesize{
\begin{tabular}{c c c}%{lcdr} 
\hline \hline 
State($\vert \psi \rangle$)/Process($\Lambda$) &~~~
$\mathcal{F}(\rho^{{\rm DQST}}_{{\rm weak}})$~~~&~~~
$\mathcal{F}(\rho^{{\rm true}}_{{\rm weak}})$~~~\\
\hline \hline 
$\vert \psi_1 \rangle =(\vert 00 \rangle + \vert 11 \rangle)/\sqrt{2} $ & $0.9511 \pm 0.0065$ &  $0.9791$   \\ 
$\vert \psi_2 \rangle =(\vert 01 \rangle + \vert 10 \rangle)/\sqrt{2} $ & $0.9266 \pm 0.0075$ &  $0.9739$  \\ 
$\Lambda_1 = H $ & $0.9447 \pm 0.0060$ &  $0.9703$   \\ 
$\Lambda_2 = R_x(\frac{\pi}{2}) $ & $0.9476 \pm 0.0029$ & $0.9729$   \\ 

\hline \end{tabular} }
\label{table3}
\end{table}
%%%%%%%%%%%%%%%%%%%%%%%%%%%%%%%%%%%%%%%%

The NMR readouts demonstrating the DQST of the Bell state $\vert \psi_1 \rangle
= (\vert 00 \rangle + \vert 11 \rangle)/\sqrt{2}$ are shown in
Fig.~\ref{bell_spectra}, where weak measurements of four Pauli operators were
carried out. The NMR readouts corresponding to the weak measurement of
$\sigma_{1z}$, $\sigma_{2x}$, $\sigma_{1x}$ and $\sigma_{1x}\sigma_{2x}$ are
depicted in red, green, purple and blue respectively, while the spectrum in
black represents the reference spectrum.  It can be clearly seen that the
non-zero spectral intensities of the NMR peaks corresponding to $(\vert 00
\rangle, \sigma_{1z})$, $(\vert 11 \rangle, \sigma_{1z})$, $(\vert 11 \rangle,
\sigma_{1x}\sigma_{2x})$ yield the three density matrix elements $\rho_{11}$,
$\rho_{44}$ and $\rho_{14}$ respectively, whereas other peak intensities
(Eq.~(\ref{rho})) tend to zero as compared to the reference spectrum. The
experimentally obtained real and imaginary parts of the density matrix
corresponding to the Bell state $\vert \psi_1 \rangle $ are given in
Eqs.~(\ref{bell_re})-(\ref{bell_im}), respectively. All the elements were
measured with considerably high accuracy and precision.

It is to be noted that the experimental density matrix is Hermitian by
construction (the imaginary part of all diagonal elements can be ignored
and set to zero) but may not satisfy positivity and trace conditions
as all the independent elements $\lbrace \rho_{ij}, i \leq j \rbrace$ are
computed individually and independently. 
For DQST of the Bell state
$\vert \psi_1 \rangle$, the trace turns out to be $1.3435$ and 
the eigenvalues are
$1.2836$, $0.2825$, $-0.1553$ and $-0.0673$, which do not
correspond to a valid density matrix. However, the true
quantum state satisfying all the properties of a valid density matrix can be
recovered from the experimental density matrix by recasting it as a constrained
convex optimization problem~\cite{gaikwad-qip-2021}:
\begin{subequations} \begin{alignat}{2} &\!\min_{\overrightarrow{\rho}^{\rm
true}_{\rm weak}}        &\qquad& \Vert \overrightarrow{\rho}^{\rm true}_{\rm
weak}-\overrightarrow{\rho}^{\rm dqst}_{\rm weak}\Vert_{l_2}\label{e12}\\
&\text{subject to} &      & \rho^{\rm true}_{\rm weak} \geq
0,\label{e12:constraint1}\\ &                  &      & { Tr}(\rho^{\rm
true}_{\rm weak}) =1.\label{e12:constraint2} 
\end{alignat} 
\end{subequations}
where ${\rho}^{\rm true}_{\rm weak}$ is the variable density
matrix
corresponding to the true quantum state to be reconstructed, while ${\rho}^{\rm
dqst}_{\rm weak}$ is the experimentally obtained density matrix using the weak
measurement-based DQST scheme. The $\rightarrow$ arrow 
denotes the vectorized form
of the corresponding matrix and $\Vert .  \Vert_{l_2}$ represents $l_2$ norm,
also known as the Euclidean norm of a vector.  The valid density matrix
${\rho}^{\rm true}_{\rm weak}$ representing the true quantum state was recovered
from ${\rho}^{\rm dqst}_{\rm weak}$  and is given in Eq.~(\ref{bell_true}).  We
note here in passing that the experimentally obtained density matrices
${\rho}^{\rm dqst}_{\rm weak}$ (or ${\rho}^{\rm true}_{\rm weak}$) corresponding
to the states $\vert \psi_1 \rangle$ and $\vert \psi_2 \rangle$ can be
interpreted as the Choi-Jamiolkowski state corresponding to the identity gate
($\Lambda = I$) and the bit flip gate ($\Lambda = \sigma_x$), respectively.

%%%%%%%%%%%%%%%%%%%%%%%%%%%%%%%%%%%%%%%%%%%%%%%%%%
\begin{widetext}
\begin{equation} \label{bell_re}
\rm{Re}(\rho^{\rm dqst}_{\rm weak})=\begin{pmatrix}
 0.6089\pm 0.0015 & -0.0104\pm 0.0007 & 0.0774\pm 0.0047 & 0.6314\pm 0.0074 \\
 -0.0104\pm 0.0007 & 0.0155\pm 0.0059 & 0.0633\pm 0.0073 & -0.0534\pm 0.0183 \\
 0.0774\pm 0.0047 & 0.0633\pm 0.0073 & 0.0764\pm 0.0019 & -0.0956\pm 0.0292 \\
 0.6314\pm 0.0074 & -0.0534\pm 0.0183 & -0.0956\pm 0.0292 & 0.6425\pm 0.0011 \\
\end{pmatrix}
\end{equation}
\begin{equation} \label{bell_im}
\rm{Im}(\rho^{\rm dqst}_{\rm weak})=\begin{pmatrix}
 0 & -0.0566\pm 0.0085 & 0.0352\pm 0.0123 & 0.1126\pm 0.0374 \\
 0.0566\pm 0.0085 & 0 & 0.0860\pm 0.0217 & -0.1384\pm 0.0036 \\
 -0.0352\pm 0.0123 & -0.0860\pm 0.0217 & 0 & 0.1367\pm 0.0139 \\
- 0.1126\pm 0.0374 & 0.1384\pm 0.0036 & 0.1367\pm 0.0139 & 0 \\
\end{pmatrix}
\end{equation}
\begin{equation} \label{bell_true}
\rho^{\rm true}_{\rm weak}=\begin{pmatrix}
 0.4667 & -0.0300 - 0.0333i & -0.0217 - 0.0618i  & 0.4858 - 0.0811i \\
  -0.0300 + 0.0333i &  0.0043  &  0.0058 + 0.0024i & -0.0255 + 0.0399i \\
  -0.0217 + 0.0618i  &  0.0058 - 0.0024i &  0.0092  &  -0.0118 + 0.0681i \\
   0.4858 + 0.0811i &  -0.0255 - 0.0399i &  -0.0118 - 0.0681i  &  0.5198  \\
\end{pmatrix}
\end{equation}
\end{widetext}
%%%%%%%%%%%%%%%%%%%%%%%%%%%%%%%%%%%%%%%%%%%%%%%%%%%%%

%%%%%%%%%%%%%%%%%%%%%%%%%%%%%%%%%%%%%%%%%%%%%%%%%
\begin{figure}[t]
\centering
{\includegraphics[scale=0.92]{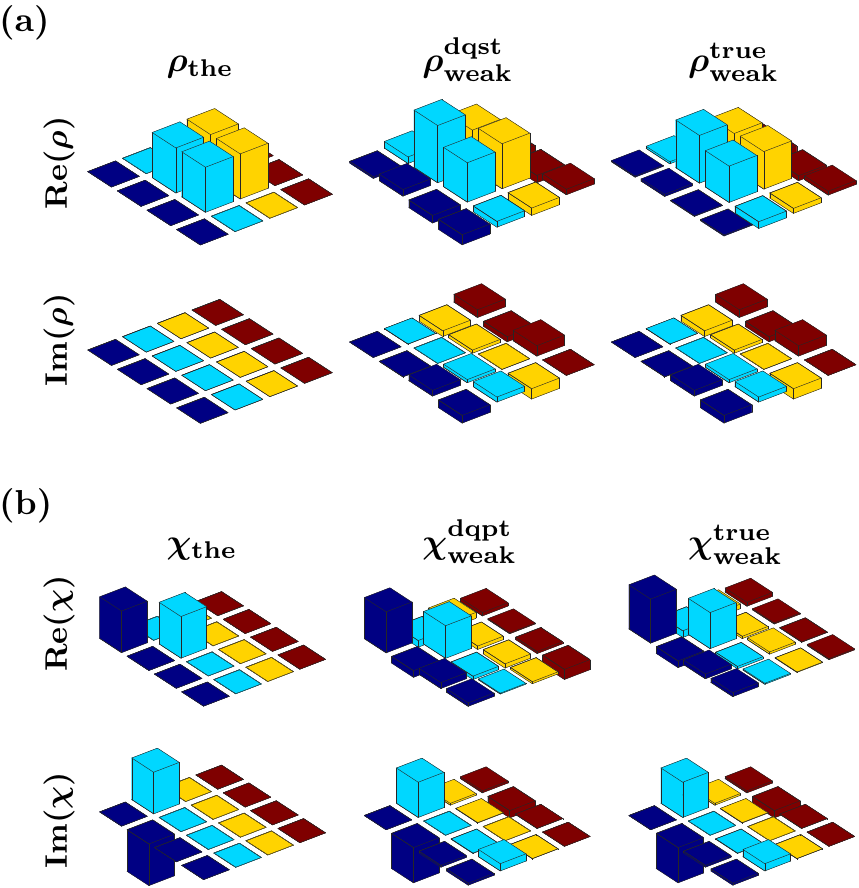}}
\caption{(Color online) Theoretical and 
experimentally reconstructed density matrices corresponding to 
(a)
the quantum state $\vert \psi_2 \rangle = 
(\vert 01 \rangle + \vert 10 \rangle)/\sqrt{2}$
and  (b) the rotation operation $R_{x}(\frac{\pi}{2})$.}
\label{tomo}
\end{figure}
%%%%%%%%%%%%%%%%%%%%%%%%%%%%%%%%%%%%%%%%%%%%%%
For DQPT implementation, the $U_{\chi}$ acts as 
a change of basis operation, which transforms 
the Choi-Jamiolkowski state to the process matrix $\chi$ in 
the chosen basis. The desired unitary operator $U_{\chi}$ is set to:
%%%%%%%%%%%%%%%%%%%%%%%%%%%%%%
\begin{equation} \label{uchi}	
U_{\chi}=\frac{1}{\sqrt{2}}
\begin{pmatrix}
1 & 0 & 0 & 1 \\
0 & 1 & 1 & 0 \\
0 & -i & i & 0 \\
1 & 0 & 0 & -1
\end{pmatrix}
\end{equation}
which allows the estimation of the process matrix $\chi$ in the Pauli basis.
The real and imaginary parts of the process matrix $\chi^{\rm dqst}_{\rm weak}$
in the Pauli basis corresponding to the  Hadamard gate $H$ obtained via the weak
measurement-based DQPT protocol are given in Eqs.~(\ref{had_re})-(\ref{had_im}),
respectively, where the trace turned out to be 0.9589 and the eigenvalues are
-0.1646, 0.0795, 0.1427 and 0.9014. The true quantum process $\chi^{\rm
true}_{\rm weak}$ can be recovered from $\chi^{\rm dqst}_{\rm weak}$ by solving
a similar convex optimization problem as given in Eq.~(\ref{e12}) with the
additional constraint $\sum_{m, n} \chi_{m n} E_n^{\dagger} E_m= I$,  and is
given in Eq.~(\ref{had_true}). The theoretical process matrix corresponding to
the Hadamard gate contains only four non-zero elements $\lbrace \rho_{ij} = 0.5
\vert i,j=2,4 \rbrace$, and it can be seen from
Eqs.~(\ref{had_re})-(\ref{had_im}) that the weak measuremement scheme is able to
determine all these elements with very high accuracy.  The theoretical and
experimental density and process matrices corresponding to the quantum state
$\vert \psi_2 \rangle$ and the gate $R_x(\frac{\pi}{2})$ are graphically
represented in Fig.\ref{tomo}.  The experimental state (process) fidelity
$\mathcal{F}$ is computed using the normalized trace distance between the
experimental and theoretical density (process) matrices\cite{harpreet-mle}. The
experimental fidelity of various quantum states and processes obtained via the
weak measurement-based protocol is given in Table~\ref{table3}.

\begin{widetext}
\begin{equation} \label{had_re}
\rm{Re}(\chi^{\rm dqst}_{\rm weak})=\begin{pmatrix}
 -0.0010 \pm 0.0005 &0.0422\pm 0.0041&0.0635\pm 0.0012&-0.0854\pm 0.0004\\
0.0422\pm 0.0041&0.3964 \pm 0.0099&-0.0827\pm 0.0004 &0.4406 \pm 0.0097\\
0.0635\pm 0.0012&-0.08269\pm 0.0004&0.0789 \pm 0.0036&0.0429\pm 0.0019\\
-0.0854\pm 0.0004&0.4406\pm 0.0097 &0.0429\pm 0.0019&0.4846\pm 0.0035\\
\end{pmatrix}
\end{equation}
  
  \begin{equation} \label{had_im}
\rm{Im}(\chi^{\rm dqst}_{\rm weak})=\begin{pmatrix}
 0&0.0243\pm 0.0033&0.0554\pm 0.0155 &-0.0195\pm 0.0012\\
-0.0243\pm 0.0033&0&-0.0664\pm 0.0081&0.0754\pm 0.0445\\
-0.0554\pm 0.0155&0.0664\pm 0.0081&0&0.0633\pm 0.0445\\
0.0195\pm 0.0012&-0.0754\pm 0.0045&-0.0633\pm 0.0004&0\\
\end{pmatrix}
\end{equation}

 \begin{equation} \label{had_true}
\chi^{\rm true}_{\rm weak}=\begin{pmatrix}
 0.0319  &  -0.0145 + 0.0072i  & 0.0144 + 0.0246i  & -0.0389 + 0.0077i \\
  -0.0145 - 0.0072i  &  0.4021  &  -0.0380 - 0.0542i   & 0.3992 + 0.0653i \\
   0.0144 - 0.0246i  & -0.0380 + 0.0542i  &  0.0831  &   0.0008 + 0.0646i \\
  -0.0389 - 0.0077i  &  0.3992 - 0.0653i  &  0.0008 - 0.0646i  &  0.4829  \\
\end{pmatrix}
\end{equation}
  
  \end{widetext}
  
\subsubsection*{Extension to $n$ qubits}
For an $n$-qubit density (or process) matrix, all the independent elements
$\lbrace \rho_{ij}, i \leq j \rbrace$ can be obtained as given in
Eq.~(\ref{rho}). All the $2^n$ diagonal elements can be recovered using a weak
measurement of the $\sigma_{1z} = \sigma_z \otimes I^{\otimes n-1}$ operator,
whereas all the $2^{n-1}(2^n-1)$ off-diagonal elements can be obtained via weak
measurements of $n$-qubit Pauli operators of the form $\lbrace I, \sigma_x
\rbrace^{\otimes n}$ (excluding $I^{\otimes n}$), each yielding $2^{n-1}$
elements.  For instance, the operator $\sigma_x^{\otimes n}$ will measure the
off-diagonal elements $\lbrace \rho_{ij}, 1 \leq i \leq 2^{n-1}, j= 2^n+1-i
\rbrace$.  The reconstruction of the full density matrix requires weak
measurements of $2^n$ Pauli operators which is in stark contrast to standard
tomographic protocols which require the measurement of $4^n-1$ operators. Hence,
even for full reconstruction, the weak measurement-based tomography protocol
turns out to be much more efficient than standard and selective tomography
protocols.  The quantum circuit given in Fig.~\ref{ckt}(a) can be extended to
$n$-qubits, with DQST requiring one extra qubit as the meter qubit and DQPT
requiring $n$ extra ancillary qubits along the meter qubit.
%%%%%%%%%%%%%%%%%%%%%%%%%%%%%%%%%%%%%%%%%
\section{Conclusions}
\label{sec4}
In this work
an efficient scheme was proposed
and  a
generalized quantum circuit to perform direct QST and QPT
using a weak measurement-based technique was
constructed and the protocol was successfully
tested  on an NMR quantum processor.  We used
the scalar $J$ coupling to control the strength of the
interaction between the system and the metre qubits and hence
were able to efficiently simulate the weak measurement
process with high accuracy.  Our protocol allows us to
directly obtain multiple selective elements of the density
and the process matrix of an unknown quantum state and an
unknown quantum process in a single experiment, which makes
it  more attractive as compared to other direct
tomography methods.  Furthermore we employed the convex
optimization method to recover the underlying true quantum
states and processes from the experimental data sets obtained
via the  weak measurement-based scheme  which substantially
improved the experimental fidelities.

Unlike other measurement-based DQST (or DQPT) methods which
require projective measurements on the system qubits and
maximally disturb the state of the system, our protocol does
not involve any  measurements on the system qubits.  Our
experiments open up new research directions for various
interesting weak measurement experiments on quantum
ensembles which were earlier not possible.
%%%%%%%%%%%%%%%%%%%%%%%%%%%%%%%%%%%%%%%
\begin{acknowledgments}
All experiments were performed on a Bruker Avance-III 400
MHz FT-NMR spectrometer at the NMR Research Facility at
IISER Mohali. 
Arvind acknowledges funding from 
the Department of Science and Technology (DST),
India, under Grant No DST/ICPS/QuST/Theme-1/2019/Q-68. 
K.D. acknowledges funding from 
the Department of Science and Technology (DST),
India, under Grant No DST/ICPS/QuST/Theme-2/2019/Q-74. 
\end{acknowledgments}

%\bibliographystyle{apsrev4-1}
%\bibliography{weaktomo}

\end{document}